\documentclass[11pt]{article}

\usepackage{url,hyperref}
\usepackage[total={6.5in,8.75in}, top=1.6in, left=0.9in, includefoot]{geometry}
\usepackage{array}

\newcommand{\memberof}{\,{\in}\,}

\newcommand{\product}[2]{ {#1} {\times} {#2} }

\newcommand{\triple}[3]{\mbox{$ \langle #1,#2,#3 \rangle $}}
\newcommand{\quadruple}[4]{\mbox{$ \langle #1,#2,#3,#4 \rangle $}}

\title{{\bf Digital Libraries, Conceptual Knowledge Systems, and the Nebula Interface}}
\author{
	
	{\bf Robert E. Kent}\thanks{
	This research was funded by a grant from {\bf Intel} Corporation.} \\
	University of Arkansas \\
	rekent@logos.ualr.edu\\  
\and
	{\bf C. Mic Bowman}\thanks{
	This research was funded by the Advanced Research Project Agency, under contract number F19628-93-C-1076.} \\
	Transarc Corporation\\
	mic@transarc.com \\
}

\date{}

\begin{document}
	\maketitle

\begin{abstract}
\noindent
Concept Analysis \cite{Wi82} provides a principled approach to effective management of wide area information systems,
such as the Nebula File System and Interface \cite{BDBCP94}.
This not only offers evidence to support the assertion \cite{MiDo94}
that a digital library is a bounded collection of incommensurate information sources in a logical space,
but also sheds light on techniques for collaboration 
through coordinated access to the shared organization of knowledge \cite{Fu94}.
\end{abstract}

\newpage

\section*{Introduction}

In their lead-off paper for last year's
{\sl Digital Library' 94\/} conference \cite{MiDo94},
Francis Miksa and Philip Doty
ask the question
``Why should a digital library be called a `library'?'' 
They examine three aspects of the traditional library
which may reveal the meaning of a digital library.
These three aspects were highlighted in their statement (our emphases) that
\begin{center}
{\footnotesize\begin{tabular}{c}
	``a library is 
	a {\em collection\/} 
	of {\em information sources\/} 
	in a {\em place\/}.''
\end{tabular}}
\end{center}
\begin{itemize}
	\item A {\em collection\/} consists of objects gathered and assembled together
		with boundaries based upon pragmatic considerations and purpose.
	\item {\em Information sources\/} are separate and unique 
		intellectual and artistic entities,
		which inform the user by highly incommensurate
		methods, purposes, and intensions.
	\item A {\em place\/} is an intellectual construct,
		a logical or intellectual space,
		with location of place meaning
		``a rationalized set of relationships''
		which structure the members of a collection.
\end{itemize}
In the announcement for this year's
{\sl Digital Library' 95\/} conference \cite{Fu94},
Richard Furuta
directs our attention to a further collaborative aspect of digital libraries,
which was highlighted in his statement (our emphases) that
\begin{center}
{\footnotesize\begin{tabular}{p{3.9in}}
``scholarly work in the digital library of the future will be mediated
through coordinated access to {\em shared information spaces\/}. 
Patrons will {\em organize\/}
their own private digital libraries, 
{\em collaborate\/} with colleagues through shared digital libraries, 
and 
have access to huge amounts of multimedia information 
in global, public digital libraries''.
\end{tabular}}
\end{center}
In this paper we describe 
a mathematical framework,
Concept Analysis and conceptual knowledge systems,
and 
an implementation prototype
in the areas of resource discovery and wide area information management,
the Nebula File System and Interface,
which potentially satisfy all four of these criteria.
\begin{itemize}
	\item The bounded collection 
		is represented by
		the notions of {\em many-valued context\/} and {\em formal context\/} in Concept Analysis
		or {\em context\/} in Nebula.
	\item The information source
		is represented by
		an {\em object\/} in Concept Analysis
		or a {\em file object\/} in Nebula.
		The incommensurability of information sources
		is represented by 
		the {\em typing\/} of synoptic information
		which has been abstracted from objects,
		and afterward is associated with them.
	\item The place, or logical space, 
		is represented by 
		the notion of a {\em concept lattice\/} in Concept Analysis
		or a {\em view taxonomy\/} in Nebula.
	\item Collaboration through coordinated access to the organization of shared information spaces,
		is provided in Nebula by the specification of connectivity 
		between information spaces via scoping,
		and is represented in Concept Analysis
		in terms of elaboration of 
		the notion of a {\em conceptual knowledge system\/}
		as a constrained sum of formal contexts.
\end{itemize}
Table~\ref{aspects} summarizes the analogies 
between these four aspects of Digital Libraries
and various notions in Concept Analysis and the Nebula Interface.
\begin{table}[htb]
\begin{center}
\setlength{\extrarowheight}{1.0pt}
{\footnotesize\begin{tabular}{|c|c|c|}\hline
	{\bf Digital Libraries} & {\bf Concept Analysis} & {\bf Nebula Interface} \\ \hline\hline
	collection            & formal      & context \\
	(boundaries)          & context     & \\ \hline
	incommensurate        & typed       & typed \\
	information sources   & objects     & file objects \\ \hline
	place                 & concept     & view \\
	(logical space)       & lattice     & taxonomy \\ \hline
	coordinated access    & conceptual  & connectivity \\
	to                    & knowledge   & specification \\
	information spaces    & system      & via scoping \\ \hline
\end{tabular}}
\end{center}
\caption{{\bf Analogies}}
\label{aspects}
\end{table}
Since libraries function as shared repositories of information,
their contents must be classified and catalogued for easy access by patrons.
Classification is a common technique for organizing information \cite{Ro87}.
A taxonomy or classification scheme is 
an orderly arrangement of items into classes
according to shared characteristics \cite{WyTa92}.
There is a strong and useful analogy 
displayed in Table~\ref{ls:rd:analogies}
between classification in Library Science
and classification in Network Information Discovery and Retrieval (NIDR) systems
such as Nebula.
The Nebula File System \cite{BDBCP94} and Interface \cite{Po94}
is a prototype wide area information system \cite{BoSpSp94},
which offers a new model 
for the organization and coordinated access of information spaces.
The Nebula system integrates
information management with traditional file system operations.
In Nebula,
resource meta-information is classified by collocation in views.
Views are conceptual classes which organize resource meta-information
into dynamically customizable information spaces.

Concept Analysis \cite{Wi82} provides a principled approach to 
the effective management of wide area information systems
such as Nebula.
It is a new approach to the analysis of information,
which
provides the mathematical foundation for faceted, synthetic classification.
This is the appropriate model for 
the organization of knowledge in dynamic view-oriented NIDR systems.
Conceptual scaling \cite{GaWi89}
provides the mathematical foundation for faceted analysis
via the user's view and interpretation of information.
A flexible dynamic organizing/browsing mechanism,
based on ideas from conceptual scaling,
is important as publishing moves 
from a ``push'' model,
where an editor determines what readers see,
to a ``pull'' model,
where users 
decide for themselves what to read
by selecting from a variety of information sources.
Conceptual knowledge systems \cite{Wi92}
provide an adequate theory of knowledge 
consisting of:
knowledge representation,
knowledge inferencing,
knowledge acquisition,
and
knowledge communication tools.
Shared access to information spaces,
as specified by Nebula scoping,
can be formally represent by conceptual knowledge systems.

\begin{table}[htb]
\begin{center}
\setlength{\extrarowheight}{1.0pt}
{\footnotesize\begin{tabular}{|c|c|}\hline
	{\bf Library Science}
		& {\bf Resource Discovery}
			\\ \hline\hline
	bibliographic description 
		& summarization \& synopsis
			\\ \hline
	bibliographic tools 
		& NIDR systems
			\\ \hline
	bibliographic control
		& information management
			\\ \hline
\end{tabular}}
\end{center}
\caption{{\bf Analogies}}
\label{ls:rd:analogies}
\end{table}

The paper is organized as follows.
Section~\ref{section:abstract} discusses the nature and benefits of 
the data abstraction known as meta-information.
It is also concerned with the interpretation of such data
by faceted analysis.
Interpreted meta-information is modeled by formal contexts in Concept Analysis.
Such contexts 
correspond to the idea that 
digital libraries are ``collections'' with boundaries.
In Section~\ref{section:organize} the organization of knowledge through synthetic classification
is described in terms of 
the notion of conceptual knowledge systems from Concept Analysis.
An instance of conceptual knowledge systems called view taxonomy
is constructed in Nebula 
by the specification of views as descriptive names within scoped indexes.
The conceptual space of such organized knowledge 
corresponds to the notion of 
digital libraries as collections of information resources ``in a place.''
Section~\ref{section:manage} is concerned with the management of information.
It discusses the twin paradigms of searching and browsing,
and introduces the idea of concept neighborhood navigation.
Section~\ref{section:share} indicates how collaboration
between digital libraries,
small local conceptual spaces (private or group)
and 
large global conceptual spaces,
can be defined via shared organization.
Finally,
Section~\ref{section:future}
summarizes what we have accomplished in this paper
and
briefly describes some future work.

\section{Abstracting Meta-Information}\label{section:abstract}

In the universe of all knowledge,
both verbal knowledge and the non-verbal knowledge in music and art,
there is a certain amount of summarization or synoptic knowledge 
that has been recorded \cite{WyTa92}.
This knowledge is called meta-information,
and is referred to as the bibliographic universe in Library Science.
This meta-information often consists of a set of tag/value attributes or elements.
The Nebula File System abstracts from files various elements of meta-information
such as filename, size, date created, owner, etc.
Nebula refers to such meta-information as a {\em file object\/}.
A file object consists of a collection of attributes
which the represented file possesses or has.
Nebula file objects are gathered together into contexts.
A Nebula context stores not only file objects,
but also views and a collection of binary index relations
that connect attribute values to file objects.
There is one index relation for each attribute tag.
Table~\ref{document:mvcxt} displays a Nebula context of documents 
(see Figure 2 in \cite{BDBCP94}).

Nebula contexts are instances
of the Concept Analysis notion of a many-valued context.
The unanalyzed and uninterpreted data of many application domains
can often be conceptualized as a constrained collection of many-valued contexts.
A {\em many-valued context\/} \cite{GaWi89}
is a quadruple $\quadruple{G}{N}{D}{\phi}$,
where:
$G$ is a set of objects,
which models the set of file objects and views in Nebula;
$N$ is a set of sorts,
which model attribute-tags in Nebula 
(and database or entity/relationship attributes);
$D = \{ D_a \mid a \memberof N \}$ 
is an $N$-indexed collection of domains of values,
corresponding to the attribute-values in Nebula; 
and
$\phi = \{ G \stackrel{\phi_a}{\rightarrow} D_a \mid a \memberof N \}$ 
is an $N$-indexed collection of functions,
called an {\em information function\/},
whose inverses correspond to the index relations in a Nebula context.
The Nebula context of documents in Table~\ref{document:mvcxt}
is a many-valued context.
The set of sorts is $N = \{ \mbox{project}, \mbox{format} \}$,
the set of objects is 
$G = \{ \mbox{plan1.ps}, \mbox{plan2.ps}, \mbox{plan2.doc}, \mbox{notes1.txt}, \mbox{notes2.txt} \}$,
the domains are
$D_{\rm project} = \{ \mbox{plan1}, \mbox{plan2} \}$
and
$D_{\rm format} = \{ \mbox{postscript}, \mbox{text} \}$, 
and
the information function
$\phi_{\rm project}(\mbox{plan1.ps}) = \mbox{plan1}$,
$\phi_{\rm project}(\mbox{notes1.txt}) = \mbox{plan2}$,
etc.

The {\bf has} relationship between Nebula file objects and attributes 
is an instance of the binary incidence matrix of a formal context.
A {\em (formal) context\/} is a triple $\triple{G}{M}{I}$
consisting of two sets $G$ and $M$
and a binary {\em incidence\/} relation $I \subseteq \product{G}{M}$ between $G$ and $M$.
Intuitively, the elements of $G$ are thought of as {\em entities\/} or {\em objects\/},
the elements of $M$ are thought of as {\em properties\/}, {\em characteristics\/} or {\em attributes\/}
that the entities might have,
and $g{I}m$ asserts that ``object $g$ {\em has\/} attribute $m$.''
Table~\ref{document:cxt} displays a formal context 
obtained by nominal conceptual scaling of the attributes
from the multi-valued context displayed in Table~\ref{document:mvcxt}.

\begin{table}[htb]
\begin{center}
\setlength{\extrarowheight}{1.0pt}
{\footnotesize\begin{tabular}{|l||l|c|} \hline
	           & \multicolumn{1}{|c|}{project} & format \\ \hline\hline
	plan1.ps   & plan1 & postscript \\
	plan2.ps   & plan2 & postscript \\
	plan2.doc  & plan2 & ---        \\
	notes1.txt & plan2 & text       \\
	notes2.txt & plan2 & text       \\
	\hline
\end{tabular}}
\end{center}
\caption{{\bf Many-Valued Context for Documents}}
\label{document:mvcxt}
\end{table}

\begin{table}[htb]
\setlength{\extrarowheight}{1.0pt}
\begin{center}
\begin{tabular}[t]{c@{\hspace{5mm}}c@{\hspace{5mm}}c}
{\footnotesize\begin{tabular}[t]{|l@{\hspace{2mm}}l|} \hline
	\multicolumn{2}{|c|}{{\bf objects}} \\ \hline\hline
	1 & plan1.ps   \\
	2 & plan2.ps   \\
	3 & plan2.doc  \\
	4 & notes1.txt \\
	5 & notes2.txt \\ \hline
\end{tabular}}
&
{\footnotesize\begin{tabular}[t]{|l|c@{\hspace{3pt}}c@{\hspace{3pt}}c@{\hspace{3pt}}c|} \hline
	\multicolumn{5}{|c|}{{\bf incidence}} \\ \hline\hline
	           & 1 & 2 & 3 & 4 \\ \hline
	1 &$\times$&&$\times$& \\
	2 &&$\times$&$\times$& \\
	3 &&$\times$&& \\
	4 &&$\times$&&$\times$ \\
	5 &&$\times$&&$\times$ \\ \hline
\end{tabular}}
&
{\footnotesize\begin{tabular}[t]{|l@{\hspace{2mm}}l|} \hline
	\multicolumn{2}{|c|}{{\bf attributes}} \\ \hline\hline
	1 & project=plan1     \\
	2 & project=plan2     \\
	3 & format=postscript \\
	4 & format=text       \\ \hline
\end{tabular}}
\end{tabular}
\end{center}
\caption{{\bf Formal Context for Documents}}
\label{document:cxt}
\end{table}

\section{Organizing Conceptual Space}\label{section:organize}

Nebula identifies resources through descriptive names \cite{Po94,BoSpSp94}.
A {\em descriptive name\/} is an expression 
which selects objects based upon their registered attributes.
A basic form of descriptive name is just a conjunction of attributes,
such as
\begin{center}\verb|format=text & project=plan2 & name=notes2.txt|.\end{center}
Descriptive names provide file name resolution through {\em associative access\/}.
They are particularly important for finding resources 
in very large information spaces when the resource location is unknown.
A set of functions for resolving descriptive names exists in each Nebula context.
These include 
attribute domain operations such as equality and order,
and regular expression matching.

Nebula replaces the directories in traditional file systems with database views.
The directory structure of traditional file systems is a static organization,
which ``reflects the requirements of the system designer,
not the varied requirements of a diverse user-base'' \cite{BDBCP94}.
Nebula views provide a powerful and flexible mechanism
for the logical and dynamic organization of an information space,
which allows user customization.
In Nebula a {\em view\/} is a query
which by using a descriptive name
selects file objects from within a scope index.
Views relate file objects by containment and scope.
Containment defines an abstraction relationship
over the file objects contain in a view.
The properties shared by those objects are abstracted in 
the descriptive name in the {\tt containment} attribute of the view.
The view serves as the conceptual class for the objects it contains.
Scope defines a generalization-specialization relationship
between the conceptual classes denoted by the views.
By using views the Nebula File System provides a classification scheme 
which is both synthetic and faceted:
synthetic, because it generates conceptual categorizations ``on the fly'';
faceted, because it uses atomic units of information 
known as facets to accomplish this categorization.

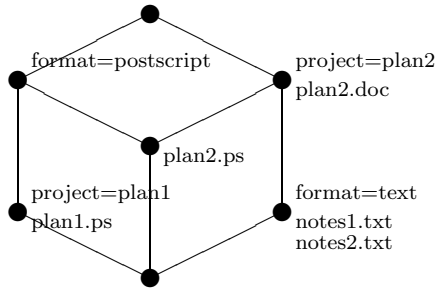
\begin{figure}[htb]
\begin{center}
\setlength{\unitlength}{1pt}
\newcommand{\putdisk}[3]{\put(#1,#2){\circle*{#3}}}
\newcommand{\puttext}[3]{\put(#1,#2){{\mbox{\scriptsize$#3$\normalsize}}}}
\begin{picture}(500,100)
\put(180,0){\begin{picture}(100,100)
	\puttext{0}{0}{{\bf }}
	\putdisk{50}{100}{7}					
	\putdisk{0}{75}{7}					
	\puttext{5}{80}{\mbox{\scriptsize\rm format{=}postscript}}
	\put(0,75){\line(2,1){50}}				
	\putdisk{100}{75}{7}					
	\puttext{105}{80}{\mbox{\scriptsize\rm project{=}plan2}}
	\puttext{105}{69}{\mbox{\scriptsize\rm plan2.doc}}			
	\put(100,75){\line(-2,1){50}}				
	\putdisk{50}{50}{7}					
	\puttext{55}{44}{\mbox{\scriptsize\rm plan2.ps}}			
	\put(50,50){\line(-2,1){50}}				
	\put(50,50){\line(2,1){50}}				
	\putdisk{0}{25}{7}					
	\puttext{5}{30}{\mbox{\scriptsize\rm project{=}plan1}}			
	\puttext{5}{19}{\mbox{\scriptsize\rm plan1.ps}}				
	\put(0,25){\line(0,1){50}}				
	\putdisk{100}{25}{7}					
	\puttext{105}{30}{\mbox{\scriptsize\rm format{=}text}}			
	\puttext{105}{19}{\mbox{\scriptsize\rm notes1.txt}}			
	\puttext{105}{11}{\mbox{\scriptsize\rm notes2.txt}}			
	\put(100,25){\line(0,1){50}}				
	\putdisk{50}{0}{7}					
	\put(50,0){\line(-2,1){50}}				
	\put(50,0){\line(0,1){50}}				
	\put(50,0){\line(2,1){50}}				
\end{picture}}
\end{picture}
\end{center}
\caption{{\bf lattice for documents}}
\label{document:lat}
\end{figure}

Nebula views are instances of 
the notion of a formal concept in Concept Analysis,
used within the confines of a conceptual knowledge system.
A {\em formal concept\/} or {\em conceptual class\/} consists of 
any group of entities or objects
exhibiting one or more common characteristics, traits or attributes.
Conceptual classes are logically characterized by their extension and intension.
The {\em extension\/} of a class is 
the aggregate of entities or objects
which it includes or denotes.
The {\em intension\/} of a class is 
the sum of its unique characteristics, traits or attributes,
which, taken together,
imply the concept signified by the conceptual class.
The process of subordination of conceptual classes and collocation of objects
exhibits a natural order,
proceeding top-down
from the more general classes with larger extension and smaller intension
to the more specialized classes with smaller extension and larger intension.
This {\bf isa} relationship is a partial order called generalization-specialization.
Conceptual classes with this generalization-specialization ordering
form a class hierarchy for the context,
which mathematically is a complete lattice,
Figure~\ref{document:lat} displays the lattice of conceptual classes
associated with the formal context of documents displayed in Table~\ref{document:cxt}.

\footnotesize
\normalsize

According to Concept Analysis,
in addition to modeling knowledge representation in file systems,
with the notion of a {\em conceptual knowledge system\/} \cite{Wi92}
we will be able to do 
knowledge inferencing, knowledge acquisition, and knowledge communication. 
There are 3 basic notions in conceptual knowledge systems: 
objects, attributes, and concepts. 
In Nebula, 
these correspond to: 
file objects, attributes, and views, 
respectively. 
There are 4 basic relationships in conceptual knowledge systems: 
an object has an attribute, 
an object belongs to a concept, 
an attribute abstracts from a concept, 
and 
a concept is a subconcept of another concept.
In Nebula, these correspond to: has, containment, constructor, and scope
relationships, respectively.
These notions and relationships partition
the frame of a conceptual knowledge system
as in Table~\ref{fca:relationships}.

\begin{table}[htb]
\begin{center}
\setlength{\extrarowheight}{1.0pt}
{\footnotesize\begin{tabular}{|r||c|c|}\hline
	incidence		& {\bf classes}	& {\bf attributes}	\\ \hline\hline
	{\bf classes}		& organization	& distinguishing	\\ \hline
	{\bf objects}		& instantiation	& having		\\ \hline
\end{tabular}}
\end{center}
\caption{{\bf Conceptual Knowledge System Relationhips}}
\label{fca:relationships}
\end{table}

\begin{figure}[htb]
\begin{center}
\setlength{\unitlength}{1pt}
\newcommand{\putcircle}[3]{\put(#1,#2){\circle{#3}}}
\newcommand{\putdisk}[3]{\put(#1,#2){\circle*{#3}}}
\newcommand{\puttext}[3]{\put(#1,#2){{\mbox{\scriptsize$#3$\normalsize}}}}
\begin{picture}(300,175)
\put(50,0){\begin{picture}(200,175)
	\puttext{0}{0}{{\bf }}
	\putdisk{100}{175}{8}					
	\puttext{105}{171}{\fbox{\tiny\rm Object}}	
	\putdisk{100}{150}{8}					
	\puttext{105}{146}{\fbox{\tiny\rm Document}}
	\put(100,150){\line(0,1){25}}			
	\putdisk{25}{100}{8}					
	\puttext{0}{108}{\mbox{\tiny\rm format{=}postscript}}
	\puttext{30}{96}{\fbox{\tiny\rm PostScript}}
	\put(25,100){\line(3,2){75}}			
	\putdisk{75}{100}{8}					
	\puttext{70}{108}{\mbox{\tiny\rm project{=}plan2}}
	\puttext{80}{96}{\fbox{\tiny\rm Plan2}}		
	\puttext{80}{85}{\mbox{\tiny\rm plan2.doc}}		
	\put(75,100){\line(1,2){25}}			
	\putdisk{125}{100}{8}					
	\puttext{125}{108}{\mbox{\tiny\rm project{=}plan1}}
	\puttext{130}{96}{\fbox{\tiny\rm Plan1}}	
	\put(125,100){\line(-1,2){25}}			
	\putdisk{175}{100}{6}					
	\puttext{180}{108}{\mbox{\tiny\rm format{=}text}}	
	\put(175,100){\line(-3,2){75}}			
	\putdisk{0}{50}{6}						
	\puttext{5}{43}{\mbox{\tiny\rm plan2.ps}}			
	\put(0,50){\line(1,2){25}}				
	\put(0,50){\line(3,2){75}}				
	\putdisk{50}{50}{6}						
	\puttext{55}{43}{\mbox{\tiny\rm plan1.ps}}		
	\put(50,50){\line(-1,2){25}}			
	\put(50,50){\line(3,2){75}}				
	\putdisk{150}{50}{6}					
	\puttext{155}{42}{\mbox{\tiny\rm notes1.txt}}		
	\puttext{155}{36}{\mbox{\tiny\rm notes2.txt}}		
	\put(150,50){\line(-3,2){75}}			
	\put(150,50){\line(1,2){25}}			
	\putdisk{200}{50}{6}					
	\puttext{205}{43}{\mbox{\tiny\rm notes0.txt}}		
	\put(200,50){\line(-3,2){75}}			
	\put(200,50){\line(-1,2){25}}			
	\putdisk{100}{0}{6}						
	\put(100,0){\line(-2,1){100}}			
	\put(100,0){\line(-1,1){50}}			
	\put(100,0){\line(1,1){50}}				
	\put(100,0){\line(2,1){100}}			
\end{picture}}
\end{picture}
\end{center}
\caption{{\bf Lattice for Document Conceptual Knowledge System}}
\label{document:cks:lat}
\end{figure}
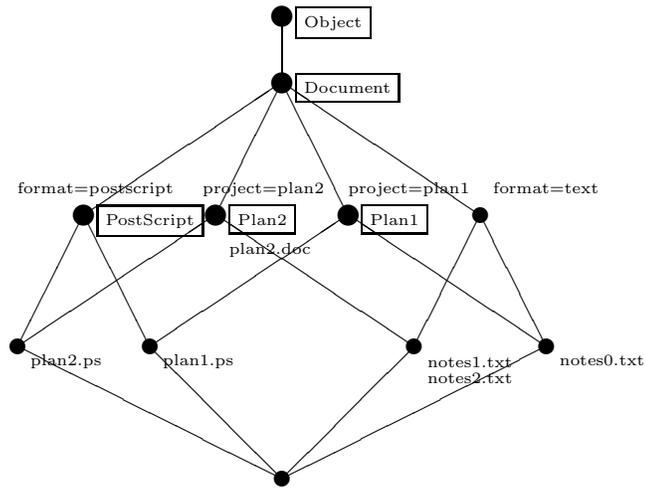

Table~\ref{document:cks:cxt} represents a conceptual knowledge system 
within the conceptual universe ${\cal D}$ 
of all documents in an information system and their properties
(see Figure 2 in \cite{BDBCP94}).
The conceptual knowledge system in Table~\ref{document:cks:cxt}
extends the formal context of documents represented by Table~\ref{document:cxt}.
It consists of a set $B$ of five concepts of ${\cal D}$ defined as views,
in addition to the set of objects and attributes in Table~\ref{document:cxt}.
For convenience of illustration,
we have added an additional object ``notes0.txt''.
The crosses in Table~\ref{document:cks:cxt} represent four relations 
(Boolean matrices):
the organization submatrix is the order relation on the 5 classes,
and the having submatrix is identical to the incidence matrix in Table~\ref{document:cxt}.
Figure~\ref{document:cks:lat} represents the lattice of conceptual classes
for the conceptual knowledge system displayed in Table~\ref{document:cks:cxt}.
The set of concepts 
$B = \{ {\rm objs}, {\rm docs}, {\rm postscript}, {\rm plan1}, {\rm plan2} \}$,
which were added to the document space
to form the conceptual knowledge system,
are each represent by a node 
at the top of the line diagram in Figure~\ref{document:cks:lat}.

\begin{table}[htb]
\begin{center}
\setlength{\extrarowheight}{1.0pt}
\begin{tabular}{c@{\hspace{5mm}}c@{\hspace{5mm}}c}
	{\footnotesize\begin{tabular}{|r@{\hspace{2mm}}l|} \hline
		\multicolumn{2}{|c|}{{\bf classes/objects}} \\ \hline\hline
		1 & Object     \\
		2 & Document   \\
		3 & PostScript \\
		4 & Plan1      \\
		5 & Plan2      \\ \hline
		6 & plan1.ps   \\
		7 & plan2.ps   \\
		8 & plan2.doc  \\
		9 & notes0.txt \\
		9 & notes1.txt \\
		10 & notes2.txt \\ \hline
	\end{tabular}}
&
{\footnotesize\begin{tabular}{|r|c@{\hspace{3pt}}c@{\hspace{3pt}}c@{\hspace{3pt}}c@{\hspace{3pt}}
	c@{\hspace{3pt}}|c@{\hspace{3pt}}c@{\hspace{3pt}}c@{\hspace{3pt}}c|} \hline
	\multicolumn{10}{|c|}{{\bf incidence closure}} \\ \hline\hline
	           & 1 & 2 & 3 & 4 & 5 & 6 & 7 & 8 & 9 \\ \hline
	1 &$\times$&&&& &&&& \\
	2 &$\times$&$\times$&&& &&&& \\
	3 &$\times$&$\times$&$\times$&& &&&$\times$& \\
	4 &$\times$&$\times$&&$\times$& &$\times$&&& \\
	5 &$\times$&$\times$&&&$\times$ &&$\times$&& \\ \hline
	6 &$\times$&$\times$&$\times$&$\times$& &$\times$&&$\times$& \\
	7 &$\times$&$\times$&$\times$&&$\times$ &&$\times$&$\times$& \\
	8 &$\times$&$\times$&&&$\times$ &&$\times$&& \\
	9  &$\times$&$\times$&&$\times$&&$\times$&&&$\times$ \\
	10 &$\times$&$\times$&&&$\times$&&$\times$&&$\times$ \\
	11 &$\times$&$\times$&&&$\times$&&$\times$&&$\times$ \\ \hline
\end{tabular}}
&
{\footnotesize\begin{tabular}{|r@{\hspace{2mm}}l|} \hline
	\multicolumn{2}{|c|}{{\bf classes/attributes}} \\ \hline\hline
	1 & Object            \\
	2 & Document          \\
	3 & PostScript        \\
	4 & Plan1             \\
	5 & Plan2             \\ \hline
	6 & project=plan1     \\
	7 & project=plan2     \\
	8 & format=postscript \\
	9 & format=text       \\ \hline
\end{tabular}}
\end{tabular}
\end{center}
\caption{{\bf Conceptual Knowledge System in the Document Universe}}
\label{document:cks:cxt}
\end{table}

\section{Managing Information}\label{section:manage}

In Library Science
only the bibliographic universe can be controlled.
Such control is performed by means of bibliographic tools.
Catalogs are bibliographic tools,
which exercise bibliographic control through three basic functions:
identification, collocation, and evaluation.
A user,
who has a citation or bibliographic item in mind,
should be able to match or {\em identify\/} or find a bibliographic entry
for that item.
This is the searching paradigm.
A user,
based upon various bibliographic data and connecting references,
should be able
to bring together in one place or {\em collocate\/}
bibliographic entries for similar and closely related material.
This is part of the browsing paradigm.
A user
should be able to choose by {\em evaluation\/} from among many bibliographic entries
the one that best represents 
the knowledge, information or specific item desired.
This also is part of the browsing paradigm.

Wide area information systems 
provide access to networked information resources
through many different application interfaces:
WWW hypertext,
Gopher menues,
and Archie search.
File management systems,
such as Archie, Gopher, Prospero, WAIS, and WWW,
use an existing file system as a storage repository.
Resource discovery systems,
such as Harvest and Whois$+$$+$,
effectively manipulate the tremendous amount of heterogeneous information
in wide area information systems and wide area file systems,
by using the uniform, logical interface provided by
the resource meta-information in bibliographic records:
Harvest defines its Summary Object Interchange Format (SOIF)
and
Whois$+$$+$ can use the resource description 
called the Uniform Resource Characteristic (URC) 
currently being developed by
the Internet Engineering Task Force (IETF) working group on
uniform resource identifiers.
Just as the ISBD in Library Science,
this meta-information is typed in order to represent 
the heterogeneity of networked information resources.
Resource discovery systems use two paradigms
for managing information: searching and organizing/browsing.

\begin{center}
{\footnotesize\begin{tabular}{p{2.2in}p{3.8in}}
{\em Searching\/} is the process of locating resources.
		The user provides a description or query of the resources being sought.
		The resource discovery system resolves the query
		and returns to the user a list of resources which match the description.
		Archie is the canonical example of 
		an information system based upon the search paradigm.
To search for a file
		the user must possess enough information about the file 
		to formulate a query.
		Searching is most effective 
		when a user can formulate a precise query
		from attributes that are indexed for efficient lookup.
		The main weakness of searching
		is that formation of good queries can be a difficult task,
		especially in an information space unfamiliar to the user.
 &
{\em Organizing\/} is the human-guided process 
		(on the server side)
		of deciding how to interrelate information,
		usually by placing it into some sort of hierarchy
		(for example, the hierarchy of direcetories in an FTP file system).
		{\em Browsing\/} is the corresponding human-guided process 
		(on the client side)
		of exploring the organization and contents of a resource space.
The main weakness of organizing/browsing
		is that 
		(1) it is done by someone else,
		(2) it is not easy to change
		(for example,
		in Library Science the standard classification systems
		of Dewey and the Library of Congress
		are fixed and possibly not relevant to the present-day patron),
		and
		(3) it is difficult to keep a large amount of data ``well organized''.
		For effective browsing
		the system does not resolve queries,
		but instead it organizes the information space
		so that the user can navigate it easily.
		The key to effective browsing
		is a well-organized, flexible, dynamic information space.
Classification is a common technique for organizing information.
		A taxonomy or classification scheme is 
		an orderly arrangement of terms or classes.
		Such a scheme arranges a set of objects 
		into classes with shared characteristics.
\end{tabular}}
\end{center}

\newpage
Concept Analysis defines a navigation method
called {\em concept neighborhood navigation\/},
which moves between local conceptual neighborhoods.
This allows for an interactive exploration of information spaces,
individual or shared.
A subset of facets can be chosen and starting from these, 
a local environment of related items can be explored. 
When browsing a very large data repository, 
it can be desirable to select a set of starting objects 
and
to interactively explore their neighborhood. 
This process consists of the following steps: 

\small
\begin{center}
\begin{minipage}{6in}
\begin{description}
\item[Initialization] \mbox{ }
The facets are evaluated 
		with respect to the uninterpreted data
		and transformed into binary relations (formal contexts).
The evaluated facets are composed into a single formal context
		using the operation of {\em apposition\/}
		(this operation requires the contexts to share a common object set).
A global analysis is performed on the total context,
		chiefly in terms of the collection of local neighborhood lattices.
\item[Browse Loop] \mbox{ }
A new seed is chosen.
		This may be either an object or an attribute.
The local neighborhood of a given seed object 
		is analyzed and previewed.
		To simplify the visualization data to be presented to the user
		(and possibly reach an acceptable number of conceptual classes),
		the local neighborhood is modified using various means:
		raising the connectivity threshold,
		rank-ordering the attributes and restricting to the most important ones,
		restricting to a ball around the seed induced by a similarity metric,
		etc.
The local neighborhood is visualized.
		At this time the user may want to visualize the union context 
		of the local neighborhoods for the old seed and the new seed
		--- this allows comparison of ``distance'' moved and classes in common.
\end{description}
\end{minipage}
\end{center}
\normalsize

\section{Sharing Organization}\label{section:share}

Scoping in Nebula
can provide coordinated access and efficient sharing 
of the organization of separate, sharable information spaces \cite{Po94}.
Comparison of the private individual information spaces of experts
in the context of psychoanalysis
has been discussed in Concept Analysis \cite{SWFL94}.
The mathematical foundations,
for collaboration by coordinated access to the knowledge organized in logical spaces,
is defined in terms of constrained sums of formal contexts \cite{Ke92}
applied to conceptual knowledge systems.
The sharing of knowledge organization can be accomplished 
with the specification of connectivity between logical information spaces.
Sharing organization between two logical spaces
is visualize in Table~\ref{sharing} 
in terms of elaborated conceptual knowledge systems.
The first logical space makes use of the organization of the second logical space 
by specifying the link connectivity $\mbox{sharing}_{1,2}$.
These links represent scoping 
the first logical space conceptual classes
on appropriate classes in the second logical space.
So with links in $\mbox{sharing}_{1,2}$
the second logical space shares its organization with the first logical space.
The linked connectivity in $\mbox{sharing}_{2,1}$ represent the dual situation
--- the sharing of the first by the second.
The cross-linked instantiation
in blocks $\mbox{instantiation}_{1,2}$ and $\mbox{instantiation}_{2,1}$
is derived connectivity defined by incidence matrix closure.

\begin{table}
\begin{center}
{\footnotesize\begin{tabular}{|r||c|c|c|}\hline
	incidence          & $\mbox{classes}_1$           & $\mbox{classes}_2$           & attributes                \\ \hline\hline
	$\mbox{classes}_1$ & $\mbox{organization}_1$      & $\mbox{sharing}_{1,2}$       & $\mbox{distinguishing}_1$ \\ \hline
	$\mbox{classes}_2$ & $\mbox{sharing}_{2,1}$       & $\mbox{organization}_2$      & $\mbox{distinguishing}_2$ \\ \hline
	$\mbox{objects}_1$ & $\mbox{instantiation}_1$     & $\mbox{instantiation}_{1,2}$ & $\mbox{having}_1$         \\ \hline
	$\mbox{objects}_2$ & $\mbox{instantiation}_{2,1}$ & $\mbox{instantiation}_2$     & $\mbox{having}_2$         \\ \hline
\end{tabular}}
\end{center}
\caption{{\bf Sharing Organization}}
\label{sharing}
\end{table}

As Francis Miksa and Philip Doty have pointed out \cite{MiDo94},
the Internet Gopher space today is not a digital library
--- it contains a huge variety of useful sources,
but these are ``not tied together as a single intellectual construct, 
neither in the sense of structure nor in the sense of access methods.''
We contend that such legacy information 
can be augmented and organized by coordinated access 
into a collection of collaborating digital libraries.
The following example,
which is part of the LC Marvel Gopher space at the Library of Congress,
describes how this can be done.

The Library of Congress 
Machine-Assisted Realization of the Virtual Electronic Library 
(LC Marvel located at \verb|gopher://marvel.loc.gov/|) 
classifies government publications.
The conceptual space defined 
includes government publications as objects 
and 
abstracted content as attributes.
Table~\ref{tab:govpub} shows attributes 
registered for three documents in the conceptual space defined by Marvel.
The conceptual space is structured 
according to government branch, office, and project.
Table~\ref{tab:govclass} shows several document classes that exist. 
When implemented in Nebula, 
each publication is a file object 
with attributes registered for relevant properties.
In practice, 
these attributes are collected by textual summarizers 
that process the typical structure of government documents 
such as press releases, newsletters, and BAA's.
A special ``class'' attribute is registered for a document
that corresponds to 
the path of Gopher menus traversed to retrieve the document.
The ``class'' attribute provides 
some additional information about the content of the document.

Nebula represents each class as a view 
that contains all objects in the corresponding menu or in any submenu.
That is, 
the ``Executive'' view contains 
all publications for the White House, 
the Department of Agriculture, 
and any other Executive branch department or committee.
In addition to the conceptual space of government publications,
each user defines a ``reader'' space 
that augments the structure of the conceptual space defined by Marvel.
The user space represents a profile of information the user finds interesting.
For example, 
a user may consider interesting 
the class of documents regarding nuclear waste disposal.
This class of documents crosses 
the boundaries of classes defined by Marvel.
In fact 
it might include 
publications by the Judicial branch,
executive orders and press releases from the White House 
and the Department of Energy, 
and legislative actions from Congress. 

Intuitively, 
the reason for constructing user classes is that 
formal, organizational classification of government publications 
is too general.
While the general classification is useful 
for browsing the collection of documents, 
it is not sufficient for issue-specific location of publications.
Nebula accommodates user classes 
by layering them on top of the more general structure.
The formal structure is available globally. 
In a separate user space, 
an issue-specific view is constructed 
by scoping it on the appropriate organizational classes. 
In this way, 
the formal structure provides a reasonable first approximation 
on which users can construct a customized structure.

\begin{table}[htb]
\begin{center}
\setlength{\extrarowheight}{1.0pt}
  {\footnotesize\begin{tabular}{lp{3.5in}}
    \hline\hline
    {\bf Field}         & {\bf Value} \\    
    \hline\hline
    title               & Department of Energy Electronic 
                          Commerce Newsletter \\
    issue               & 3 \\
    publication-date    & December 1994 \\
    published-by        & Department of Energy \\
    internet-contact    & ECNews@hq.doe.gov \\
    section-headings    & Updates from the Pilot Sites, Electronic
                          Resources, Central Registration, EC Forum \\
    \hline
    title               & Scalable Systems and Software \\
    number              & BAA-95-18 \\
    published-by        & Advanced Research Projects Agency \\
    program-manager     & Robert Parker, Glenn Ricart \\
    \hline
    title               &  Remarks by the Vice President in Swearing-In
                           Ceremony of Supreme Court Justice 
                           Stephen G. Breyer \\
    publication-date    &  August 12, 1994 \\
    published-by        &  The White House, Office of the Press
                           Secretary \\
    \hline\hline
\end{tabular}} 
\end{center}
\caption{{\bf Attributes for Three Example Publications}}
\label{tab:govpub}
\end{table}

\begin{table}[htb]
\begin{center}
\setlength{\extrarowheight}{1.0pt}
{\footnotesize\begin{tabular}{ll}
\\
    \hline\hline
    Executive                 &   Judiciary                         \\
    Executive:White House     &   Judiciary:Supreme Court           \\
    Executive:Agriculture     &   Judiciary:Supreme Court:Justices  \\
    Executive:Energy          &   Judiciary:Supreme Court:Decisions \\
    Legislative                    &   Military                     \\
    Legislative:House              &   Military:Navy                \\ 
    Legislative:Senate             &   Military:ARPA                \\
    Legislative:Technology Assesme &   Military:ARPA:Solicitations  \\
    \hline\hline
\end{tabular}}
\end{center}
\caption{{\bf Several Classes of Government Publications}}
\label{tab:govclass}
\end{table}

\section{Conclusions \& Future Work}\label{section:future}

In this paper we have demonstrated in a very real sense
how the Nebula prototype and Concept Analysis
articulate the idea of digital libraries
as
bounded collections of typed information sources in conceptual spaces
with collaboration defined by coordinated access to the shared organization of knowledge.

In particular,
we have shown how Concept Analysis provides 
a principled foundation for the Nebula Interface,
by giving it a mathematically rigorous base, 
which fits well with the intuitions behind Nebula.
For example,
the notion of a concept lattice provides an explicit mathematical structure 
for Nebula view taxonomies.
In addition,
Concept Analysis reveals the composite nature of view specification.
The first step is the explicit specification of 
the lattice join of the superordinate views in the {\tt scope} component.
The second step is 
the specialization of the superordinate join view
via the conceptually scaled attributes in the {\tt constructor} component.

The Concept Analysis foundation
should allow us to apply the same extensions to Nebula 
which have been given to Concept Analysis:
the Fuzzy Logic extension of Concept Analysis \cite{Ke94b}
will allow us to extend Nebula to fuzzy taxonomies of views;
and
the Rough Set extension of Concept Analysis \cite{Ke93b}
will allow us to define rough Nebula views.
Concept Analysis strongly suggests 
the use of faceted analysis via conceptual scaling,
along with the current awareness ideas of SDI and continuous queries,
to serve as a basis for the descriptive name component 
in the specification of views.
It suggests the importance of ``implications'' and expert system type rules
as an augmented means for analyzing taxonomic structures in Nebula.

A formal basis for describing views in terms of objects 
and a more expressive environment 
for analyzing the relationships between views 
may make it possible to simplify a collection of views,
thus providing query/space optimization. 
In the simple case, 
two views could be collapsed into one 
if they always contains the same set of objects.
This is a straightforward application of 
the Concept Analysis optimization technique of ``purification''.
Also applicable is the optimization technique of ``reduction'',
which eliminates objects and attributes which are not irreducible.

More generally, 
a ``cluster'' of similar (but not identical) views 
might be abstracted into a more general view 
that many would find interesting.
In Concept Analysis
this kind of clustering has been captured 
by the notion of a {\em rough concept\/} in the Rough Set extension, 
which is based upon an indiscernibility relation on objects 
(which could dually be on attributes).
It can also be realized by
the notion of a generalized metric over conceptual classes (views),
which would cluster views by a tolerance setting 
(a ball around a view of a certain small radius).

Traditional browsing systems,
such as Gopher and the World Wide Web,
define browsing transitions along 
the physical links of the knowledge organization,
whether these are hierarchical as in Gopher-space
or cross-referential as in the Web.
Concept Analysis offers a more flexible and logical alternative.
By introducing the idea of a concept neighborhood,
it defines browsing transitions along 
conceptual links of the knowledge organization.
The knowledge organization here consists of the whole information space,
which would normally be only virtually represented.
One travels 
from the local concept neighborhood of one object
to the local concept neighborhood of a neighboring object.
The transition can be pictured by the union neighborhood.

Future work will include:
the incorporation into Nebula of Fuzzy Logic and Rough Set extensions;
application of conceptual scaling techniques from Concept Analysis
to provide for a rigorous foundation and elaboration
of the associative access in Nebula view specification;
realization of a testbed for collaborative studies; and
development of client software for analysis and browsing 
by concept neighborhood navigation.

	\bibliography{dl95}
	\bibliographystyle{plain}

\end{document}